\def\EE#1{\times 10^{#1}}
\def\msun{\hbox{M$_{\odot}$}}
\documentclass[structabstract]{aa}  
\usepackage{graphicx}
\usepackage{txfonts}
\begin{document}
   \title{VLBI imaging of M\,81* at 43\,GHz}


   \author{Eduardo Ros
          \inst{1,2}
          \and
          Miguel \'A.  P\'erez-Torres\inst{3}
}

   \institute{Departament d'Astronomia i Astrof\'{\i}sica, Universitat
		de Val\`encia, E-46100 Burjassot, Val\`encia, Spain\\
              \email{Eduardo.Ros@uv.es}
         \and
		Max-Planck-Institut f\"ur Radioastronomie, Auf dem
		H\"ugel 69, D-53121 Bonn, Germany
         \and
		Instituto de Astrof\'{\i}sica de Andaluc\'{\i}a, 
		CSIC, Apdo. Correos 2004, E-08071, Granada, Spain\\
             \email{torres@iaa.es}
}

   \date{Submitted: \today}


 
  \abstract
{The nearby spiral galaxy \object{M\,81} harbors in its core 
  a Low-Luminosity AGN
  (LLAGN), and appears closely related to the more distant and
  powerful AGNs seen in quasars and radio galaxies.  The intrinsic
size of this object is unknown due to scattering, and 
it has shown a core-jet morphology with weak extended emission rotating
with wavelength.}
{The proximity of \object{M\,81} ($D=3.63$\,Mpc) allows a
  detailed investigation of its nucleus to be made.
  The nucleus is four orders of magnitude more
  luminous than the Galactic centre, and is therefore considered a link
  between Sgr\,A$^\star$ and the more powerful nuclei of radio
  galaxies and quasars.  {Our main goal was to determine the
    size of \object{M\,81$^*$} at a shorter wavelength thus
    directly testing whether
    the frequency-size dependent law $\Theta \propto \nu^{-0.8}$ was
    still valid for wavelengths shorter than 1\,cm. In addition, we
    also aimed to confirm the rotation of the source as a
    function of frequency.}}
{We observed the continuum 7\,mm radio emission of \object{M\,81$^*$}
  using the Very Long Baseline Array on Sep 13, 2002, using nearby
  calibrators to apply their interferometric observables to the target
  source, to increase the chances of detection.  The source was
  detected on all baselines and hybrid mapping was possible.}
{We present the first 7\,mm VLBI image of the core of {\object{M\,81$^*$}},
  which represents the highest resolution image
  ever of this source.  Modeling the interferometric visibilities with
  two Gaussian functions sets constraints on the angular size of its
  core down to 38 microarcseconds, {corresponding to a maximum
    (projected)
    linear size of 138\,AU, 
} and shows extended emission towards the NE
  with a position angle of $\sim50^\circ$.  A fit of one Gaussian
  elliptical function yields a position angle of $28\pm8$ degrees for
  its elongated, compact structure.  Combining the 7\,mm size with
earlier measurements at other frequencies we determine
   a frequency-size dependence
of $\Theta\propto\nu^{(-0.88\pm0.04)}$.
}
{{Our VLBI imaging of {\object{M\,81$^*$}} has clearly detected its
    core-jet structure, and has allowed us to 
estimate a size for its core, with a minimum size 
   of 138\,AU ($\approx$\,100~Schwarzschild radii).
    Our work opens the avenue for further observations of
    {\object{M\,81$^*$}} at high-angular resolution, including the monitoring of its
    structure, given that much higher bandwidths are
    currently available on the interferometric networks. In
    particular, this would allow testing for possible proper motions
    of the core or of its components in the inner jet of
    \object{M\,81$^*$}, as well as for the speed of the detected jet
    components.}}


   \maketitle
%

\section{Introduction}

The spiral galaxy \object{M\,81} (\object{B0951+633},
\object{J095532+69038}, \object{NGC\,3031}, \object{Z\,333$-$7},
\object{UGC\,5318}) at a distance of {$D =$3.63\,Mpc, 
(Freeman et al.\ \cite{fre94})}
harbors, together with the Seyfert~2 galaxy \object{Centaurus\,A}
($D$$\sim$3.4\,Mpc,
e.g., Ferrarese et al.\ \cite{fer07}, M\"uller et al.\ \cite{mue11}),
the closest extragalactic active nucleus.  \textit{Hubble Space
  Telescope} (\textit{HST}) spectroscopic observations imply a central 
mass of $7\EE{7}$ \msun\ (Devereux et al.\ \cite{dev03}).  The nucleus of
\object{M\,81} (hereafter referred to as
\object{M\,81$^*$}
)
emits in radio with a compact structure
  and exhibits both low-ionization nuclear emission-line region
  (LINER; Heckman \cite{hec80}) 
and Seyfert~1 characteristics. It
  thus appears closely related to the more distant and powerful
active galactic nuclei (AGNs) seen in quasars and radio galaxies.
However, \object{M\,81$^*$} 
occurs in a spiral rather than in an elliptical galaxy and is
relatively small and faint.  In fact, very-long-baseline
interferometry (VLBI) observations (e.g., Bartel et al.\ \cite{bar82})
have shown that the central source is about 1000--4000 AU across,
depending on observing frequency, and that the radio luminosity of
\object{M\,81$^*$} is $\approx$\,$10^{37.5}$\,erg\,s$^{-1}$, which
classifies \object{M\,81$^*$} as a low-luminosity AGN.  Nonetheless,
its size is more than two orders of magnitude---and its luminosity
about four orders---greater than that of the central source in our
Galaxy, Sgr A$^*$ (e.g., Reuter \& Lesch \cite{reu96}).
  Therefore, the proximity of
{\object{M\,81$^*$}} {allows for a detailed investigation of its nucleus
to study the link between}
 our (weak) Galactic centre
and the more powerful nuclei of radio galaxies and quasars.  
  {Indeed, X-ray and radio observations show that the ratio of the 
(5\,GHz) radio--to--X-ray (soft) luminosity for \object{M\,81$^*$} can
  vary from $R_{\rm X} \sim 1.8\EE{-5}$ up to values of $R_{\rm X}
  \sim 3.5\EE{-4}$, with the former value being typical of
  radio-quiet radio galaxies, and the latter of
  radio-loud LLAGN (see 
Terashima \& Wilson 2003), which
  suggests that \object{M\,81$^*$} shares properties of both kinds of
  objects, or may transit from one to another with time.}

As seen with VLBI (Bietenholz et al.\ \cite{bie00}; Mart\'{\i}-Vidal
et al.\ \cite{mar10}), the {nucleus} of {\object{M\,81$^*$}} shows a
stationary feature in its structure---tentatively identified with the
core---with a one-sided jet to the north east.  {The apparent source size
  is $\Theta \sim0.5$\,mas ($\sim$1800\,AU) at 8.4\,GHz, and follows a
  power-law with frequency ($\Theta \propto \nu^{-0.8}$) between
  2.3\,GHz and 22\,GHz.} The sky orientation of the radio structure 
of {\object{M\,81$^*$}} seems
to be also frequency dependent, changing from $\sim 75^\circ$ at
2.3\,GHz to $\sim 40^\circ$ at 22\,GHz.



Variability of \object{M\,81$^*$} has been reported in several
occassions, from X-rays, {both long-term (e.g., Ishisaki et
  al. \cite{ish96}) and short-term (e.g., Markoff et al.\
  \cite{mar08})}, {up to millimeter (Sch\"odel et al.\
  \cite{sch07}; Markoff et al.\ \cite{mar08}) and radio (e.g., Ho et
  al.\ \cite{ho99}; Mart\'{\i}-Vidal et al.\ \cite{mar10})
  wavelengths.  The turnover frequency of the synchrotron spectrum is
  in the range 150--200\,GHz (Sch\"odel et al.\ \cite{sch07}, Doi et
  al.\ \cite{doi11}).  In particular,} VLA observations by Ho et al.\
(\cite{ho99}) at wavelengths of 2\,cm to 20\,cm in the mid 1990s
showed outbursts lasting up to three months.  Mart\'{\i}-Vidal et
al. (\cite{mar10}) has {recently} shown that the peaks of flux
density are shifted with frequency when variability occurs, {which
  can be interpreted as  being due to opacity effects in the inner regions of
  the jet.}  \object{M\,81$^*$} is also very peculiar, together with
Sgr\,A$^*$, in its polarisation properties, since it shows significant
circular polarisation (Brunthaler et al.\ \cite{bru01}), rather than
being linearly polarised, which is much more common among AGNs.


{VLBI observations at frequencies above 22\,GHz were lacking for \object{M\,81$^*$}, 
limiting our ability to set even tighter constraints on the 
size of \object{M\,81$^*$}.
Even from space-VLBI observations at 5\,GHz (Bartel \& Bietenholz \cite{bar00}),
the source structure remained unresolved.
Therefore, we observed this radio source at a frequency of 43\,GHz ($\lambda$7\,mm) using
the Very Long Baseline Array (VLBA),} with the {main} goal of getting the
size of the object {at 43\,GHz. In turn, this
  would represent a direct test of the apparent frequency-size
  dependence ($\Theta \propto \nu^{-0.8}$) observed at lower
  frequencies.




\section{Observations and data reduction}
\label{obs}

We observed \object{M\,81$^*$} on September 13, 2002, using the
complete VLBA (Napier \cite{nap91}) at a frequency of 43\,GHz in
left-hand circular polarisation.  Data were recorded with 2\,bit
sampling at an aggregate data rate of 256\,Mbit\,s$^{-1}$, splitting
the band {in} 8 intermediate frequency (IF) channels of 8\,MHz each
(full bandwidth of 64\,MHz).

The Seyfert~1 Galaxies \object{3C\,147} 
and \object{3C\,286} 
were used as primary calibrators (fringe finding, 
delay offsets, etc.) of the observation.
We used as the phase-reference calibrator the BL\,Lac-type object
\object{B0954+658} (\object{J0958+6533}, $z$=0.368, 34\,arcmin apart
in the sky).  From UT\,14:43 to UT\,20:30 we cycled
between the target source
\object{M\,81$^*$} (60\,sec) and \object{B0954+658} (30\,sec).  The
source \object{B0951+699} was also used for phase-referencing
{testing} purposes during 30\,min (UT\,14:10 to UT\,14:43) as 
a phase-reference test target.

Without phase-referencing considerations, 
at the observed data rate, the baseline sensitivity of the
VLBA for an integration time of $\sim60$\,s was of 66\,mJy
(1-$\sigma$) for a sub-band of 8\,MHz as used in the observations
(23\,mJy for the whole band of 64\,MHz).  So, a source with
a total flux of $\sim100$\,mJy and resolved structure will hardly
be detected by long baseline on a sub-band.  
Assuming that the source would be detected on all VLBA baselines,
3-hr of integration would provide an image with a $1\,\sigma$ thermal noise
of 0.26\,mJy\,beam$^{-1}$.
The overall observing time was 5.75\,hr.  Data were correlated at
the Array Operations Center of the National Radio Astronomy
Observatory (NRAO) in Socorro, NM, USA.

We followed standard procedures to a first calibration of the
amplitudes and phases using $\cal AIPS$.  An {\em a priori} amplitude
calibration was performed using measured system temperatures and gain
curves.  After 2-bit-sampling digital correction, amplitude
calibration and the removal of the parallactic angle phase, a
single-band delay and phase {offsets} were calculated automatically
with the measured phase-cal values at each antenna.  The North-Liberty
VLBA station was used as reference antenna during the whole
procedure. The Brewster antenna was flagged from the data due to bad
performance.

The main process of the data reduction, the search for group delay and
phase rate calibration with the task \textsc{fring} in $\cal AIPS$,
had to be performed in several steps.  Direct \textsc{fring}ing on
\object{B0951+699}, \object{3C\,147}, and \object{3C\,286} did not
provide a satisfactory percentage of detections.  For
{\object{B0954+658}} we got in a first attempt detections more than 80\% of
the time.  The {data for this} source were then exported and
imaged in \textsc{difmap}, {where a few iterations of phase and 
amplitude self-calibration were
applied.}  The amplitude was only corrected for an overall factor for
each antenna.  This factor was applied to all sources back in $\cal
AIPS$.  The CLEAN components of the data resulting from \textsc{difmap} were also imported
back to $\cal AIPS$ and was used as input for the task \textsc{fring}
to obtain new values for the phase and rate for the calibrator source.
Those values were then transferred to the the target source
\object{M\,81$^*$}, and the 
 $\cal AIPS$ task \textsc{imagr} was applied to
these data {to obtain  a preliminary} phase-referenced image {of our target
source.}

Once we had a solution for the rates and delays for \object{M\,81$^*$}, 
transferred
from B0954+658, we narrowed the search windows in \textsc{fring} 
and performed a new search for the target source, with a signal-to-noise
satisfactory detection threshold of 3, to get a satisfactory number 
of visibilities with a high detection rate.  
The resulting data set could then be exported to \textsc{difmap} and
hybrid mapping with phase and (eventually) amplitude self-calibration was
performed {on {\object{M\,81$^*$}}}.

\section{Results and Discussion}
\label{results}

\subsection{Phase-referencing}

The phase-referenced image obtained after this process had a weak peak
with a
small offset with respect to the position used in the correlator.  The value for
the position was obtained by applying the $\cal AIPS$ task \textsc{jmfit}
to the image obtained with \textsc{imagr}.  
The \textsc{jmfit} position is therefore 
$\alpha=09^\mathrm{h}55^\mathrm{m}33\rlap{.} ^\mathrm{s}172932$ and
$\delta=69^\circ03^\prime55\rlap{.} ^{\prime\prime}0610744$.  The uncertainty
provided by \textsc{jmfit} for both values (which does not account for
systematics in the astrometric data reduction) is {0.12\,milliarcseconds}. 
This should not be compared directly with the phase-referenced positions to
\object{SN\,1993J} reported by other authors, since another reference
has been used. Different
attempts to get a phase-referenced image after removing different
antennas were done, and we noticed that the first two hours of
observations and the removal of the antennas in Hancock (system
temperature, $T_\mathrm{sys}$, values of above 150\,K and relative
humidity values, $h_r$, of $\sim40$\% were measured; the other antennas
had values of $\sim100$\,K and dryer atmospheres) and Saint-Croix
($T_\mathrm{sys}$ values above 200\,K and $h_r$ values above 60\%)
from the data set had the effect of improving the
phase-referenced image.  

	\begin{table}
	\caption{\label{table:modelfit} Gaussian model fit results}
	\[
	\centering
	\resizebox{0.99\columnwidth}{!}{%
	\begin{tabular}{@{}lc        c          c         c         c         c         c@{}}
	\hline
	\hline
	\noalign{\smallskip}
	& & & & Major & Axis & Position \\
	 Model &
	$S$ &
	$\Delta\alpha$ &
	$\Delta\delta$ &
	Axis &
	Ratio &
	Angle &
	$\chi^2$ \\
	 & 
	{[mJy]} & 
	{[mas]} & 
	{[mas]} & 
	{[mas]} & 
	&
	{[deg]} & \\
	\noalign{\smallskip}
	\hline
	\noalign{\smallskip}
	1-comp., circ. & 141 & ... & ... & 0.08 & 1 & ... & 1.31 \\
	\noalign{\smallskip}
	\hline
	\noalign{\smallskip}
	1-comp., ellip. & 144 & {...} & {...} & 0.162 & 0.286 & 28 & 1.26 \\
	\noalign{\smallskip}
	\hline
	\noalign{\smallskip}
	2-comp., circ., core & 122 & {...} & {...}  & 0.038 & 1 & ... & 1.26 \\
	2-comp., circ., jet  & 32 & 0.149 & 0.124 & 0.232 & 1 & {...} & 1.26 \\ 
	\noalign{\smallskip}
	\hline
	\noalign{\smallskip}
	2-comp., ellip., core & 138 & ... & ... & 0.140 & 0 & 38 & 1.20 \\
	2-comp., ellip., jet & 20 & 0.252 & 0.210 & 0.554 & 0 & 18 & 1.20 \\
	\noalign{\smallskip}
	\hline
	\end{tabular}
	}
	\]
	\end{table}

\subsection{Hybrid mapping}

For the data set with direct detections by \textsc{fring} after
using the calibrator's solutions (that is not the phase-referenced
solution and we preferred this approach due to the bad weather
during the observations producing high tropospheric
systematics), we could perform hybrid mapping with
standard procedures, involving multiple iterations of CLEAN and
phase self-calibration.  
The resulting {angular} resolution for the observed $(u,v)$-distribution
 time is of approximately 
0.36$\times$0.17\,mas at a P.A.\ of $\sim 0^\circ$.

\begin{figure}[t!]
\centering
\includegraphics[width=0.90\columnwidth]{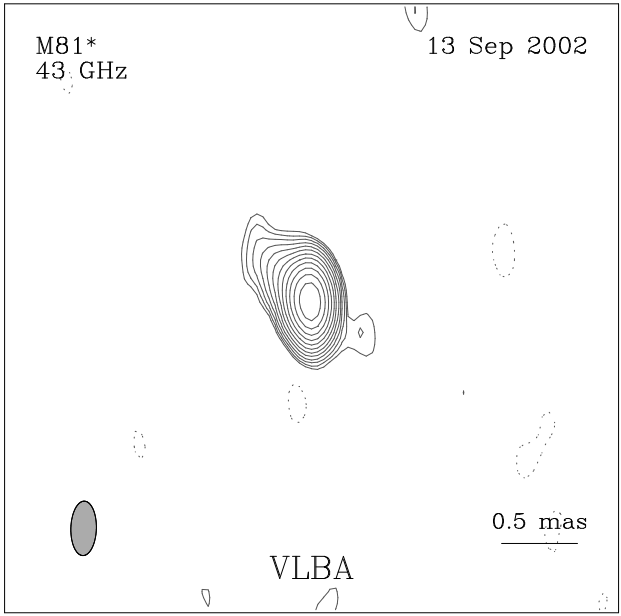}
\caption{\label{fig:m81map}
Contour image of \object{M\,81$^*$} obtained by hybrid mapping on 
the self-calibrated data set.
Contours are logarithmic, separated by a factor $\sqrt{2}$,
with the lowest level set at 1.8\,mJy/beam, that is, 
$3$\,times the root-mean-square noise of the image.
The peak of brightness is of 112\,mJy/beam.
The interferometric beam, shown at the bottom left, has a size
of $0.361\times0.168$\,mas of position angle $-1.6^\circ$.
}
\end{figure}

Amplitude self-calibration was only performed once, with a smoothing
time segment of 1\,hr.  The amplitude corrections remained {within}
15\% of the original calibration which also included the gain
corrections for the calibrator B0954+658 (see the previous Sect.).
Applying shorter time intervals to the amplitude self-calibration
did not provide satisfactory results, likely due to the large uncertainties
in the visibilities.  The resulting hybrid map, produced with natural
weighting, is shown in Fig.~\ref{fig:m81map}.  The root-mean-square
noise reached in the image was 560\,$\mu$Jy\,beam$^{-1}$.
The emission to the SW is not significant and does not affect
the model fitting results shown below.
The structure of \object{M\,81$^*$} is very compact, with hints of emission
towards the NE.  This {structure} information 
is present in the closure-phases, {and is a robust
  result}.  {Indeed, any attempt to remove this emission to the NE,
  while trying to get it in any other direction (by setting
  appropriate clean windows) or self-calibrating the data without
jet emission, yielded unsuccessful results, with the
  residual map always showing the need of emission towards the NE.}

\subsection{Gaussian model fitting}

Using the calibrated data set, we model fitted the interferometric 
visibilities with Gaussian functions to parametrize the source
emission. {We tried to reproduce  the observed visibilities by
  using {four} different fitting procedures}: 
 {(i)} a single, Gaussian, circular function (component); 
(ii) a single, elliptical Gaussian; {(iii)} two circular Gaussians, and
{(iv)} two elliptical Gaussians.  The results are shown in 
Table~\ref{table:modelfit} (flux density, offset in right ascension
and declination with respect to the main component, 
major axis of the Gaussian function, axis ratio, position angle
of the major axis, and reduced likelihood parameter).  
Our results obtained for a single, elliptical Gaussian are comparable
with earlier publications (see Bartel et al.\ \cite{bar95}).



The fit of two circular Gaussians allowed us to quantify both the core 
and jet emission.  The main, {central} component accounts for 
80\% of the whole emission, {while the remaining $\sim$20\% is in
  the  jet component, which} has a position 
angle of $\sim50^\circ$.


%
%
\begin{table}
\caption{\label{table:m81sizes}\object{M\,81$^*$} VLBI Sizes}
\[
\centering
\resizebox{0.99\columnwidth}{!}{%
\begin{tabular}{@{}c c r@{$\pm$}l r@{$\pm$}l r@{$\pm$}l r@{$\pm$}l l@{}}
 \hline \hline
\noalign{\smallskip}
 & &
\multicolumn{2}{c}{Major} & 
\multicolumn{2}{c}{Axis} & 
\multicolumn{2}{c}{Position} & 
\multicolumn{2}{c}{Equivalent} & 
\\
Frequency 	& 
Epoch(s) & 
\multicolumn{2}{c}{Axis} & 
\multicolumn{2}{c}{Ratio} & 
\multicolumn{2}{c}{Angle} & 
\multicolumn{2}{c}{Diameter} & 
Ref. \\
(GHz) 	& & 
\multicolumn{2}{c}{(mas)} 
& 
\multicolumn{2}{c}{~} 
&
\multicolumn{2}{c}{(deg)} 
& 
\multicolumn{2}{c}{(mas)} 
& 
\\
\noalign{\smallskip}
\hline
\noalign{\smallskip}
1.7 	& 1998--2005 	& 2.24&0.12 	& 0.44&0.08 	& 65&5 		& 1.48&0.29 	& 1 \\ 
2.3 	& Mar 1981 	&  1.2&0.1 	& 0.5&0.1 	& 75&3 		& 0.8&0.4 	& 2 \\ 
2.3 	& 1995--2000 	&  1.6&0.4 	& 0.26&0.24 	& 59&4 		& 0.78&0.78 	& 1 \\ 
5 	& Jun 1993 	&  0.70&0.12 	& \multicolumn{2}{c}{0} 
							& 62&8 		& \multicolumn{2}{c}{...} 
											&  3 \\ 
5   	& Apr 1976 	& \multicolumn{2}{c}{$\leq 0.4$} 
					& \multicolumn{2}{c}{1} 
							&  \multicolumn{2}{c}{...} 
									& \multicolumn{2}{c}{$\leq 0.4$} 
											& 4 \\ 
5 	& 2001--2005 	& 0.9&0.2 	& 0.48&0.41 	& 61.6&1.7	& 0.6&0.3 	& 1 \\ 
8.3 	& Mar 1981 	& 0.50&0.04 	& 0.68&0.07 	& 50&6  	& 0.41&0.06  	& 2 \\ 
8.4 	& 1993--1996 	& 0.53&0.10 	& 0.34&0.07 	& 50&4 		& 0.31&0.09 	& 5 \\ 
8.4 	& Jun 2005 	& 0.65&0.07 	& 0.9&0.2 	& 81&10 	& 0.61&0.09 	& 6 \\ 
8.4 	& 1993--2000 	& 0.49&0.15 	& 0.3&0.2 	& 51.1&1.5 	& 0.28&0.12 	& 1 \\ 
8.4 	& May/Jun 1993 	& 0.45&0.05 	& 0.47&0.18 	& 57&5 		& 0.31&0.09 	& 3 \\ 
14.9 	& Jun 1993 	& 0.31&0.04 	& 0.5&0.3 	& 48&7 		& 0.22&0.09 	& 3 \\ 
22.2 	& Apr/May 1993 	& 0.17&0.03 	& 0.35&0.18 	& 40&8 		& 0.10&0.04 	& 3 \\ 
\textbf{43} 
	& \textbf{Sep 2002} 
			& \textbf{0.16}&\textbf{0.03} 
					& \textbf{0.29}&\textbf{0.16} 
							& \textbf{28}&\textbf{8} 
									& \textbf{0.086}&\textbf{0.013} 
											& \textbf{7} \\ 
\noalign{\smallskip}
\hline
\noalign{\smallskip}
\multicolumn{11}{@{}l@{}}{\footnotesize \textbf{Note:} The values for each bibliographical reference have been averaged in frequency.}\\
\multicolumn{11}{@{}l@{}}{\footnotesize \textbf{References:}
(1) 
Mart\'{\i}-Vidal et al.\ (2010);
(2) 
Bartel et al.\ (1982); 
(3)
Bietenholz et al.\ (1996); 
} \\
\multicolumn{11}{@{}l@{}}{\footnotesize 
(4)
Kellermann et al.\ (1976); 
(5)
Bietenholz et al.\ (2000); 
(6)
Markoff et al.\ (2008); 
} \\
\multicolumn{11}{@{}l@{}}{\footnotesize 
\textbf{(7) this work}.
}
\end{tabular}
}
\]
\end{table}

\begin{figure}[t!]
\centering
\includegraphics[width=0.80\columnwidth]{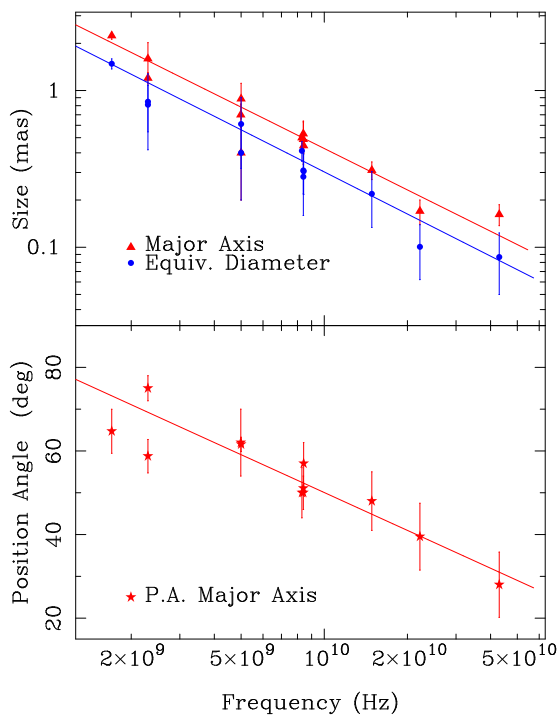}
\caption{\label{fig:m81size+pa}
Size \textbf{(top)} and position angle \textbf{(bottom)} 
of \object{M\,81$^*$} as a function of frequency.  We plot the
data from Table~2.  The two lines in the top panel are power-laws 
fit to the shown
data.  The slopes are, respectively, 
$-0.88\pm0.04$, and
$-0.89\pm0.08$, for
the major axis and the equivalent diameters of the source.
%
The line in the bottom panel, drawn as a
guide, is the function 
$\phi=(350\pm36)^\circ-(30\pm4)^\circ\times(\log(\nu/\mathrm{Hz}))$.
We did not use the data from Ref.\ 6 in Table~2, since this is the
core component only, and an extended jet component, 1\,mas away, 
was also fitted
(in this way, is not comparable with the other values, which fit
core and jet together).
}
\end{figure}


{Table~\ref{table:m81sizes} shows
earlier published data for \object{M\,81$^*$} at
different observing frequencies together with our own data (last row).
The table contains the relevant parameters characterizing the source
structure: major axis, axis ratio, position angle, and equivalent
diameter.
{The equivalent diameter is a measure of the
area covered by the Gaussian function.}
Our size determination for  \object{M\,81$^*$}  is given in the last
row of the table, and corresponds to the best-fit to a single,
elliptical Gaussian component (see Table 1). 
The uncertainties shown for the size determination at 43 GHz have been calculated from
the response of $\chi^2$ to small variations (in small steps of 1\%
for the major axis and axis ratio up to $\pm10\%$ and of $1^\circ$ in
a range $\pm10^\circ$ for the position angle of the major axis).

Figure~\ref{fig:m81size+pa}~(top) shows the size of  \object{M\,81$^*$} as a function of frequency, using
two different parameters: (i) the size of the major axis of the source
(red triangles), and (ii) the equivalent diameter of the source (blue
circles). }
A power law fit to the major axis {results in} a dependence of size
$\propto \nu^{-0.88\pm0.04}$.  The same fit to the equivalent diameter
yields essentially the same dependence, size $\propto \nu^{-0.89\pm0.08}$.
{The model predictions at 43\,GHz for the two
parameters are 
$\sim0.12$\,mas 
and $\sim0.082$\,mas, 
respectively, which are} smaller than the measured quantities of $0.160\pm0.025$\,mas, and
$0.087\pm0.037$\,mas, respectively (even with the very large
uncertainties).  If we compare the value for the size of
the nucleus as taken from the one-component elliptical Gaussian
fit (162\,$\mu$as in the longest dimension, with an equivalent
diameter of 86\,$\mu$as) with the one obtained
from the model fit including two circular Gaussian components, from
the brightest one (38\,$\mu$as) we see a margin for uncertainty in 
the fit.  In any case, the comparison with the longer wavelengths is valid
and we have found limits to the size of the emitting region in 
\object{M\,81$^*$}
at $\lambda$\,7\,mm.  

Therefore, the value found for the angular size of  \object{M\,81$^*$} is 
in the
range (162--45)\,$\mu$as (major and minor axis, respectively) down
to a lower limit of 38\,$\mu$as.  Those values correspond to
linear sizes of (590--163)\,AU to 138\,AU, respectively.
Notice that for the central mass of $7\EE7$\,\msun, the Schwarzschild
radius corresponds to 1.4\,AU. 
This is the most tight constraint on the size of \object{M\,81$^*$} ever. 
We also note that size of the major axis follows very well a power-law
with frequency, although our measurement may suggest a flattening of
this trend, following the same method as earlier authors.
The core brightness temperatures determined from the model fit sizes in
Table~\ref{table:m81sizes} are in the range $10^{(10.1-10.8)}$, which is
of the same order of magnitude as the values for LLAGNs reported
in Anderson \& Ulvestad (\cite{and05}), and is therefore below the
inverse Compton limit.

If we plot the position angle of the major angle of the ellipse as a
function of the logarithm of frequency (Fig.~\ref{fig:m81size+pa}, top), a clear
linear trend is evident, with our data point confirming previous
results reported by Bartel, Bietenholz, \& Rupen (\cite{bar95}), among other
authors.  A logarithmic fit to the data points yields a negative trend
of P.A.$=(350\pm36)^\circ-(30\pm4)^\circ \log(\nu/\textrm{Hz})$,
implying that the source orientation rotates northwards with
increasing frequency.  Following this trend, the P.A.\ of the core
at the turnover frequency of $\approx$\,200\,GHz would be 
$\sim$\,10$^\circ$, this value being the most probable.  
Several scenarios can explain this apparent rotation: a strong jet 
bending as suggested by Mart\'{\i}-Vidal et al.\ (\cite{mar10}) model fitting;
a wide angle opening in the base of the jet, visible at high frequencies;
or even if the 'core' corresponds to radio emission of the accretion
disk region.  Future observations at high frequencies with astrometric
registration of the core region should give the answer.

\section{Summary}

\label{conclusions}

We have presented the highest resolution image ever of the nucleus of
{\object{M\,81$^*$}}, and have set up a stringent constraint on the
(projected) size of its core of 138\,AU, or $\sim$100
Schwarzschild radii. 
By making use of  existing size VLBI measurements for
\object{M\,81$^*$} from 1.7 up to 22.2 GHz, and adding up our
measurement at 43 GHz, we find that the size of the core and jet 
region of {\object{M\,81$^*$}} is
best-fit by a frequency-size dependent power-law 
$\propto \nu^{-0.84\pm0.04}$, in agreement with previous results. 
Our 43 GHz data point may suggest, though, a flattening of the
power-law at frequencies around, or above, 43 GHz.

Our work opens an avenue for future, multi-epoch high-resolution VLBI
observations at 43 GHz, as well as at higher frequencies, which will
help elucidate some of the most important issues yet unsolved for
\object{M\,81$^*$}. In particular, such observations would be very
useful in constraining parameters of the radio emission models for
\object{M\,81$^*$}, as well as providing valuable information
regarding the structure variability of the core-jet, e.g., the
lifetime of the perturbations traveling down the jet.



%

\begin{acknowledgements}
We acknowledge J.\ Anderson for careful reading and 
very useful comments to the manuscript.
The 
Very Long Baseline Array is operated
by the National Radio Astronomy Observatory, a facility of the National
Science Foundation operated under cooperative agreement by Associated
Universities, Inc.
This research has made use of NASA's Astrophysics Data System.
E.R. acknowledges
partial support by the Spanish MICINN through grant
AYA2009-13036-C02-02, and by the COST action MP0905 ``Black Holes in a
Violent Universe".
M.A.P.T. acknowledges partial support by the Spanish MICINN through
grant AYA2009-13036-C02-01, {co-funded with FEDER funds,} and by
the Consejer\'{\i}a de Innovaci\'on, Ciencia y Empresa of the Junta de
Andaluc\'{\i}a through grants FQM-1747 and TIC-126. 
\end{acknowledgements}


\begin{thebibliography}{}
\bibitem[2005]{and05} Anderson, J., \& Ulvestad, J.~U. 2005, \apj, 627, 674 
\bibitem[1982]{bar82} Bartel, N., Shapiro, I.~I., Corey, B.~E., et al. 1982, \apj, 262, 556
\bibitem[1995]{bar95} Bartel, N., Bietenhoz, M.~F., \& Rupen, M.~P. 1995, Proc.\ Natnl.\ Acad.\ Sci., 92, 11374 
\bibitem[2001]{bar00} Bartel, N. \& Bietenholz, M.~F. 2000, in Astrophysical Phenomena revealed by Space VLBI, ed.\ H. Hirabayashi, P.~G. Edwards, \& D.~.W. Murphy, (Sagamihara, Japan: Institute of Space and Astronautical Science), pp. 17--20 
\bibitem[1996]{bie96} Bietenholz, M.~F., Bartel, N., Rupen, M.~P., et al. 1996, \apj, 457, 604
\bibitem[2000]{bie00} Bietenholz, M.~F., Bartel, N. \& Rupen, M.~P. 2000, \apj, 532, 895
\bibitem[2001]{bru01} Brunthaler, A., Bower, G.~C., Falcke, H., \& Mellon, R.~R. 2001, \apj, 560, L123
\bibitem[2003]{dev03} Devereux, N., Ford, H., Tsvetanov, Z., \& Jacoby, G. 2003, \aj, 125, 1226 
\bibitem[2011]{doi11} Doi, A., Nakanishi, K., Nagai, H., Kohno, K., \& Kameno, S. 2011, \aj, in press (\texttt{arXiv:1106.5627})
\bibitem[2007]{fer07} Ferrarese, L., Mould, J.~R., Stetson, P.~B., et al. 2007, \apj, 654, 186
\bibitem[1994]{fre94} Freedman, W.~L., Hughes, S.~M., Madore, B.~F., et al. 1994, \apj, 427, 628
\bibitem[1980]{hec80} Heckman, T.~M. 1980, \aap, 87, 152
\bibitem[1999]{ho99} Ho, L.~C., et al. 1999, \aj, 118, 118 
\bibitem[1996]{ish96} Ishisaki, Y., Makishima, K., Iyomoto, N., et al. 1996, \pasj, 48, 237 
\bibitem[1976]{kel76} Kellermann, K.~I., Shaffer, D.~B., Pauliny-Toth, I.~I.~K., Preuss, E., \& Witzel, A. 1976, \apj, 210, L121
\bibitem[2008]{mar08} Markoff, S., Nowak, M., Young, A., et al. 2008, \apj, 681, 905 
\bibitem[2010]{mar10} Mart\'{\i}-Vidal, I., Marcaide, J. M., Alberdi, A., et al. 2010,  \aap, 533, A111
\bibitem[2011]{mue11} M\"uller, C., Kadler, M., Ojha, R., et al. 2011, \aap, 530, L11 
\bibitem[1991]{nap91} Napier, P.~J. 2001, in ASP Conf. Ser. Vol. 19, IAU Coll.\ 131, Radio Interferometry: Theory, Techniques and Applications, ed.\ Cornwell T.~J.\ \& Perley R.~A.\ (San Francisco: ASP), pp. 390--394
\bibitem[1996]{reu96} Reuter, H.-P., \& Lesch, H. 1996, \aap, 310, L5 
\bibitem[2007]{sch07} Sch\"odel, R., Krips, M., Markoff, S., Neri, R., \& Eckart, A. 1997, \aap, 463, 551 

\end{thebibliography}
\end{document}